\documentclass[preprint,preprintnumbers,prl,amsmath,amssymb,nofootinbib,tightenlines]{revtex4}
\usepackage{graphicx}
\usepackage{dcolumn}
\usepackage{bm}
\usepackage{slashed}
\begin{document}

\title{Radiative $\beta$ Decay for Studies of CP Violation}

\author{Susan Gardner and Daheng He}

\affiliation{Department of Physics and Astronomy, University of Kentucky, 
Lexington, KY 40506-0055 
}


\begin{abstract}
A triple-product correlation in the radiative
$\beta$ decay rate of neutrons or of nuclei, characterized by the kinematical variable
$\xi\equiv (\mathbf{l}_\nu\times\mathbf{l}_e)\cdot\mathbf{k}$, 
where, e.g., $n(p) \rightarrow p(p') + e^-(l_e) + \overline{\nu}_e(l_\nu) + \gamma(k)$, 
can be generated by the pseudo-Chern-Simons term found by 
Harvey, Hill, and Hill as a consequence of the baryon vector current
anomaly and SU(2)$_{L}$$\times$U(1)$_Y$ gauge invariance at low energies. 
The correlation probes the imaginary part of its coupling constant, 
so that its observation at anticipated levels of sensitivity would reflect the
presence of sources of CP violation beyond the standard model. 
We compute the size of the asymmetry in $n\to p e^- \bar\nu_e \gamma$ decay 
in chiral effective theory, 
compare it with the computed background
from standard-model final-state interactions, and 
consider the new physics scenarios which would be limited by its experimental study. 
\end{abstract}


\maketitle

{\it Introduction ---} 
The first B-factory era, with key input from the Tevatron, established that both  
CP and flavor violation in flavor-changing processes are dominated by the 
Cabibbo-Kobayashi-Maskawa (CKM) mechanism~\cite{Isidori:2010kg}. 
The CKM mechanism, however, cannot explain the observed value of the 
baryon asymmetry of the universe, so that the problem of the missing antimatter
still weighs upon us. A path to its resolution could lie in the discovery of
non-zero values for observables which are inaccessibly small if calculated in the 
standard model (SM). Permanent electric dipole moments (EDMs) 
of nondegenerate systems, which violate T and P, are specific examples of such 
``null'' tests~\cite{pospritz}. In this paper we consider a different sort of null 
test, the pseudo-T-odd correlations of $\beta$ decay, so-called because they can 
only be motion-reversal odd~\cite{sachs}. Consequently they can be mimicked
by CP-conserving final-state interactions (FSI) in the SM, though these can be computed. 

The triple-product correlations observable in ordinary neutron or 
nuclear $\beta$ decay are
all T violating in that they are motion-reversal odd 
and connect, through an assumption of CPT
invariance, to constraints 
on sources of CP violation beyond the standard model (BSM). 
They are also spin dependent. 
In this context the study of radiative $\beta$ decay opens a new possibility,
namely, of constructing a triple-product correlation from momenta alone.
Consequently its measurement would constrain new spin-independent
sources of CP violation. Harvey, Hill, and Hill have found that interaction vertices 
involving the nucleon $N$, photon $\gamma$,  and weak gauge bosons at low energies 
emerge from gauging the axial anomaly of QCD under the full electroweak symmetry 
of the SM~\cite{HHHprl,HHHprd}. 
Such interactions can yield a triple-product momentum
correlation in the radiative $\beta$ decay of neutrons and nuclei. 
The correlation is both P and pseudo-T-odd, and 
it vanishes in the SM 
save for effects induced by FSI.
Nevertheless, the correlation can 
be generated by sources of CP violation BSM, 
and such couplings, being spin-independent, 
are not constrained by 
the nonobservation of permanent EDMs. 
We discuss, in turn, the physical origins of 
a triple momentum correlation in the decay rate, 
its possible size in different systems, and its comparison to 
the asymmetry induced by 
electromagnetic FSI in the 
SM. 

{\it Anomalous Interactions at Low Energies ---}
Radiative corrections in gauge theories need not respect all the symmetries present
in a massless Dirac theory; in particular, the axial vector current is no longer 
conserved 
and becomes anomalous. This physics is also manifest in effective theories of QCD at low
energies, in which the pseudoscalar mesons, 
interpreted as the Nambu-Goldstone bosons of a spontaneously broken chiral symmetry, 
are the natural degrees of freedom. In this context the 
nonconservation of the axial current is captured through 
the inclusion of the Wess-Zumino-Witten (WZW) term~\cite{wess,witten}, so that 
the chiral Lagrangian can then describe processes such as 
$K\bar K \to 3\pi$ and $\pi^0 \to \gamma\gamma$~\cite{witten2}. 
If we study the gauge invariance of the WZW term 
in vector-like gauge theories such as QED, 
then the vector current is conserved~\cite{adlerbardeen}. 
Harvey, Hill, and Hill have observed, however, that the gauging of this term 
under the full electroweak gauge group 
SU(2)$_L\times$U(1)$_Y$ makes the baryon vector current
anomalous and gives rise to pseudo-Chern-Simons contact 
interactions, containing $\varepsilon^{\mu\nu\rho \sigma}$, 
at low energy~\cite{HHHprl,HHHprd}. 
Such structures are also found in a chiral effective theory in terms of 
nucleons, pions, and a complete set of SM electroweak gauge fields, where the 
impact of the use of the full electroweak gauge structure of the SM 
is illuminated through use of the limit in which the Higgs vacuum expectation
value $v_{\rm weak} \gg f_\pi$ 
with the SU(2)$_L$ coupling $g_2$ small~\cite{Hill2010}.  
Thus the $W^\pm$ and $Z$ appear explicitly in the low-energy effective theory,
and the requisite terms appear at N$^2$LO in the chiral expansion. 
Namely, 
\begin{equation}
{\cal L}^{(3)} = ... + 
\frac{c_5}{M^2} \bar N i\varepsilon^{\mu\nu\rho\sigma}\gamma_\sigma \tau^a
{\rm Tr}(\tau^a \{\tilde A_\mu,[i\tilde D_\nu,i\tilde D_\rho]\}) N + ... \,,
\label{c5}
\end{equation}
where we report the charged-current term only and note 
$\tilde A_\mu$ is a SU(2) matrix of axial-vector gauge fields, 
$\tilde D_\mu$ is a covariant derivative which contains
a SU(2) matrix of vector gauge fields, $N$ is a 
nucleon doublet, and $M$ is nominally the nucleon mass. 
We refer to Ref.~\cite{Hill2010} for all details. Restoring the $W^\pm$ mass to 
its physical value, we remove the $W^\pm$ from the effective theory 
to find for neutron beta decay, e.g., 
\begin{equation}
- \frac{4 c_5}{M^2} \frac{e G_F V_{ud}}{\sqrt{2} }
\varepsilon^{\sigma\mu\nu\rho} \bar p \gamma_\sigma n 
\bar \psi_{e L} \gamma_\mu \psi_{\nu_e L} F_{\nu\rho} \,,
\end{equation}
where $2 \psi_{e L} = (1-\gamma_5) \psi_e$ and
$F_{\nu\rho}$ is the electromagnetic field strength tensor. 
Thus the baryon weak vector current can mediate parity violation on its own, 
through the interference of the leading vector amplitude mediated by 
\begin{equation}
\frac{G_F V_{ud}}{\sqrt{2}}
g_V \bar p \gamma^\mu  n \bar \psi_{e} \gamma_\mu (1 - \gamma_5) \psi_{\nu_e}\,,
\end{equation} 
dressed by bremsstrahlung from the charged particles, 
with the $c_5$ term. An analogous interference term is possible in 
neutral weak current processes. 
The T-odd momentum correlation probes the imaginary part of $g_V c_5$ 
interference. Existing constraints on $c_5$ are poor and come directly
only from the measured branching ratio in neutron 
radiative $\beta$ decay~\cite{rdk,rdk10}, as we shall consider explicitly. 
The best constraint on 
Im $g_V$ comes from the recent $D$ term measurement~\cite{nistD,nistD12}, to yield 
Im $g_V < 7\times 10^{-4}$ at 68\% CL~\cite{nistD12}. Thus a first limit 
on Im($g_V c_5$) would limit Im($c_5$). A triple-product momentum correlation
is also possible in theories BSM which do not strictly obey the $V-A$ law; 
we recall the general parametrization of Lee and Yang~\cite{leeyang56}, updated
to use the metric and conventions of Ref.~\cite{bjd} 
using Ref.~\cite{serot86}: 
\begin{eqnarray}
\nonumber H_{\mathrm{int}}& = & (\overline{\psi}_p\psi_n)(C_S\overline{\psi}_e\psi_\nu-C'_S\overline{\psi}_e\gamma_5\psi_\nu)+ 
(\overline{\psi}_p\gamma^\mu\psi_n)(C_V\overline{\psi}_e\gamma_\mu\psi_\nu-C'_V\overline{\psi}_e\gamma_\mu\gamma_5\psi_\nu) \\
\nonumber && 
+ (\overline{\psi}_p\gamma_5\psi_n)(C_P\overline{\psi}_e\gamma_5\psi_\nu-C'_P\overline{\psi}_e\psi_\nu)
- 
(\overline{\psi}_p\gamma^\mu\gamma_5\psi_n)(C_A\overline{\psi}_e\gamma_\mu\gamma_5\psi_\nu-C'_A\overline{\psi}_e\gamma_\mu\psi_\nu) \\
 && 
+ \frac{1}{2}(\overline{\psi}_p\sigma^{\mu\nu}\psi_n)(C_T\overline{\psi}_e\sigma_{\mu\nu}\psi_\nu-C'_T\overline{\psi}_e\sigma_{\mu\nu}\gamma_5\psi_\nu) \,,
\label{ly56}
\end{eqnarray}  
where in the SM $C_V=C_V'\ne 0$, $C_A=C_A'\ne 0$, 
and all other $C_i^{(')}$ vanish. There 
is an one-to-one map between these coefficients 
and those derived using modern 
effective field theory techniques at leading power in the new-physics scale, 
incorporating the exact gauge symmetry of the SM~\cite{ciri}. If the operators 
are dressed by bremsstrahlung from the charged particles, they 
can also contribute to radiative $\beta$ decay and generate a triple-product momentum
correlation. 

{\it T-odd Correlation in Radiative $\beta$ Decay --- }
In $n(p_n) \rightarrow p(p_p) + e^-(l_e) + \overline{\nu}_e(l_\nu) +
\gamma(k)$ decay the interference of the $c_5$ term 
with the leading $V-A$ terms~\cite{gaponov,bgmz,svgdh} yields the following contribution 
to the decay rate  
\begin{equation} 
|{\cal M}|^2_{c_5} = 256 e^2 G_F^2 |V_{ud}|^2
\hbox{Im}\,(c_5\,g_V) \frac{E_e}{l_e\cdot k} 
(\mathbf{l}_e\times\mathbf{k})\cdot\mathbf{l_\nu} + \dots \,,
\label{toddc5}
\end{equation} 
where we neglect corrections of radiative and recoil order. 
The pseudo-T-odd interference term is finite as $\omega \equiv k^0 \to 0$, so 
that its appearance is compatible with Low's theorem~\cite{low}. 
Alternatively, if we employ Eq.~(\ref{ly56}), we find 
\begin{eqnarray}
|{\cal M}|^2_{\mathrm{T-odd, LY}}= 16 e^2G^2_F|V_{ud}|^2 M\mathbf{l}_\nu\cdot(\mathbf{l}_e\times \mathbf{k})\frac{1}{l_e\cdot k}\mathrm{Im}[\tilde C_T(\tilde C'^\ast_S+\tilde C'^\ast_P)+
\tilde C'_T(\tilde C^\ast_S+\tilde C^\ast_P)] 
\label{toddLY}
\end{eqnarray} 
to leading radiative and recoil order, 
noting 
 $C_i^{(')}\equiv G_F V_{ud}\tilde C_i^{(')}/\sqrt{2}$ for 
 convenience; our result is compatible with that of Braguta et al.~\cite{bragutaBSM} 
in kaon radiative $\beta$ decay, $K^+ \to \pi^0 l^+ \nu \gamma$, though
they have employed a less general effective Hamiltonian. 

\begin{table}[t]
\caption {T-odd asymmetries arising from Eq.~(\ref{toddc5}), 
in units of $\mathrm{Im}\,{\cal C}_{\rm HHH}\,[\mathrm{MeV}^{-2}]$,  
for neutron, $^{19}\mathrm{Ne}$, 
and $^{35}\mathrm{Ar}$ radiative beta decay as a function of 
the minimum photon energy $\omega_{\rm min}$. 
The branching ratios are reported as well. 
}
\label{HHHdat}
\centering   \footnotesize
\vskip 0.5\baselineskip
\begin{center}
\begin{tabular}{l*{16}{c}r}
$\omega_{\rm min}({\rm MeV})$ 
&& ${\cal A}^{\mathrm{HHH}}(n)$ && $\mathrm{BR}(n)$ && ${\cal A}^{\mathrm{HHH}}(^{19}\mathrm{Ne})$ && $\mathrm{BR}(^{19}\mathrm{Ne})$ && ${\cal A}^{\mathrm{HHH}}(^{35}\mathrm{Ar})$ 
&& $\mathrm{BR}(^{35}\mathrm{Ar})$\\
\hline
0.01 && $-5.61\times10^{-3}$  
&& $3.45\times10^{-3}$  &&
$-3.60\times10^{-2}$ && $4.82\times10^{-2}$ && -0.280
&& 0.0655\\
0.05 && $-1.30\times10^{-2}$  && $1.41\times10^{-3}$  &&
$-6.13\times10^{-2}$ && $2.82\times10^{-2}$ && -0.431 && 
0.0424 \\
0.1 && $-2.20\times10^{-2}$ && $7.19\times10^{-4}$   
  && $-8.46\times10^{-2}$ && $2.01\times10^{-2}$ && -0.556
&& 0.0328\\
0.3 && $-5.34\times10^{-2}$  && $8.60\times10^{-5}$ 
&& -0.165 && $8.86\times10^{-3}$ && -0.943 && 0.0185\\
\hline
\end{tabular}
\end{center}
\vspace{1ex} 
\end{table} 
Defining $\xi\equiv (\mathbf{l}_e\times\mathbf{k})\cdot\mathbf{l}_\nu$,  
we partition phase space into regions of definite sign, so that 
we form an asymmetry:
\begin{equation} 
\mathcal{A}(\omega_{\rm min})
 \equiv 
\frac{{\Gamma_{+}(\omega_{\rm min}) - 
\Gamma_{-}(\omega_{\rm min})}}{{\Gamma_{+}(\omega_{\rm min}) + \Gamma_{-}(\omega_{\rm min})}} \,,
\end{equation} 
where $\Gamma_\pm$ contains an integral of the spin-averaged $|{\cal M}|^2$ over
the region of phase space with $\xi \stackrel{>}{{}_{<}} 0$, respectively, neglecting
corrections of recoil order. 
We compute the branching ratio (BR) as a function of
$\omega^{\rm min}$, the minimum detectable photon energy, 
ignoring terms of ${\cal O}(c_5^2)$, as well as the BSM
contributions of Eq.~(\ref{ly56}), and employing the 
inputs of Ref.~\cite{pdg12}, noting $e^2=4\pi \alpha$ with 
$\alpha\approx 1/137$ the fine-structure constant. 
As examples of nuclear radiative $\beta$ decays, 
we consider 
$^{19}\mathrm{Ne}\rightarrow ^{19}\mathrm{F} + e^+ + \nu_e + \gamma$
and 
$^{35}\mathrm{Ar}\rightarrow ^{35}\mathrm{Cl} + e^+ + \nu_e + \gamma$, namely, 
decays involving nuclear mirror transitions. 
In our evaluations, we employ the 
nuclear masses of Ref.~\cite{masstbl}, noting that the 
maximum positron energy, which is determined 
in leading recoil order by the nuclear mass difference $Q_{EC}$, is 
$3.23883\pm 0.00030$ MeV for $^{19}\mathrm{Ne}$ decay and  
$5.96614\pm 0.00070$ MeV for $^{35}\mathrm{Ar}$ decay, and the half-lives of
the compilation of Ref.~\cite{severijnsft}, 
noting 
$t_{1/2} [^{19}\mathrm{Ne}]=17.248\pm 0.029\,{\rm s}$ 
and $t_{1/2} [^{35}\mathrm{Ar}]=1.7752\pm 0.0010\,{\rm s}$. 
The asymmetries are also sensitive to 
the Gamow-Teller to Fermi mixing parameter, $\rho$~\cite{severijnsft}, 
which can be determined from either the measured decay 
rates~\cite{babrown,brownWild,severijnsft}
or the measured decay correlations~\cite{calaprice2,NaviliatCuncic:2008xt} 
in these $\beta$ decays. 
The $\rho$ values from the two methods are in agreement, except for the most
recent $^{19}\mathrm{Ne}$ results, which are 
only marginally so. The first method is more precise --- 
we use the $\rho$ values of Ref.~\cite{severijnsft}, namely, 
$\rho [^{19}\mathrm{Ne}] = -1.5933 \pm 0.0030$ and 
 $\rho [^{35}\mathrm{Ar}] = 0.2841 \pm 0.0025$ as per 
the conventions of Eq.~(\ref{ly56}). 
For the neutron we note $\rho/\sqrt{3}=\lambda= -1.2701$~\cite{pdg12}.
In Table~\ref{HHHdat}, we display these results and 
the asymmetries associated with Eq.~(\ref{toddc5}), reported 
in units of $\mathrm{Im}\,{\cal C}_{\rm HHH}\equiv \mathrm{Im}[g_V (c_5/M^2)]$, 
where we refer to Ref.~\cite{svgdh} for all details.  All nuclear calculations
are in the impulse approximation computed in leading recoil order. 
As for the asymmetry 
${\cal A}^{\rm LY}$, Eq.~(\ref{toddLY}) shows that 
the contribution exists only at second order in the recoil expansion.
In specific, 
noting $\mathrm{Im}\,{\cal C}_{\rm LY}\equiv 
\mathrm{Im}[\tilde C_T(\tilde C'^\ast_S+\tilde C'^\ast_P)
+\tilde C'_T(\tilde C^\ast_S+\tilde C^\ast_P)]$, we have 
for $\omega^{\rm min}=0.3\,{\rm MeV}$, in units of 
$\mathrm{Im}\,{\cal C}_{\rm LY}$
\begin{equation}
{\cal A}^{\rm LY}(n) =5.21\times 10^{-6}\, 
\quad ; \quad
{\cal A}^{\rm LY}(^{19}{\rm Ne}) =4.53\times 10^{-7}\, 
\quad ; \quad
{\cal A}^{\rm LY}(^{35}{\rm Ar}) =8.63\times 10^{-7}\, 
\end{equation}
All the asymmetries grow 
larger as $\omega^{\rm min}$ increases. The asymmetries associated
with Eqs.~(\ref{toddc5}) and (\ref{toddLY}) appear of grossly dissimilar size; 
however, if the 
$M$ associated with 
$\mathrm{Im}\,(c_5/M^2)$ is set by the nucleon mass, as in the SM, then
${\cal A}^{\rm HHH}$ is suppressed significantly --- though we already know the asymmetry
vanishes in the SM. Since 
the experimental figure of merit is determined by ${\cal A}^2 \mathrm{BR}$, the use of 
larger values of $\omega^{\rm min}$ would be more suitable for empirical studies. 
Moreover, the analytic structure of Eqs.~(\ref{toddc5}) and (\ref{toddLY}) 
show that the T-odd correlation also increases if the energy 
released in the decay increases. 
We discuss the criteria for choosing optimal nuclear systems later. 

We consider existing empirical constraints on the coefficients of 
Eqs.~(\ref{toddc5}) and (\ref{toddLY}). 
As for $\mathrm{Im}\,(c_5/M^2)$, the best and perhaps only 
constraint comes from the precision measurement of the branching ratio 
of neutron radiative $\beta$ decay, which has a contribution which goes as
$|c_5|^2$. We note $|{\hbox{Im}}(c_5/M^2)| < 12\, \hbox{MeV}^{-2}$ at 68\% CL
from the most recent measurement of the branching ratio for neutron 
radiative $\beta$ decay~\cite{rdk10}, for which $\omega \in [15,340]\,$keV. 
The constraint is poor because 
the radiative decay rate is driven by the contributions from the lowest photon energies,
for which $|{\cal M}|^2$ is proportional to $\omega^{-2}$~\cite{bgmz}. 
If one could measure the photon energy spectrum, e.g., close to its endpoint, 
then the constraint could be much stronger. That is, in the event that one
could measure the BR to within 1\% of its SM value for $\omega_{\rm min}=100\,$keV,
or for $\omega_{\rm min}\approx\omega_{\rm max}=782\,$keV, one would find at 68\% CL
the limits 
$|{\hbox{Im}}(c_5/M^2)| < 0.88\, \hbox{MeV}^{-2}$ and 
$|{\hbox{Im}}(c_5/M^2)| < 0.15\, \hbox{MeV}^{-2}$, respectively. 
In constrast the empirical limits on the couplings which appear in 
Eq.~(\ref{toddLY}) are already sufficiently severe~\cite{Severijns,bhattacharya}, 
that any measured asymmetry would necessarily be attributed to the 
coefficients of Eq.~(\ref{toddc5}). 

\begin{figure}
\includegraphics[width=0.9\textwidth]{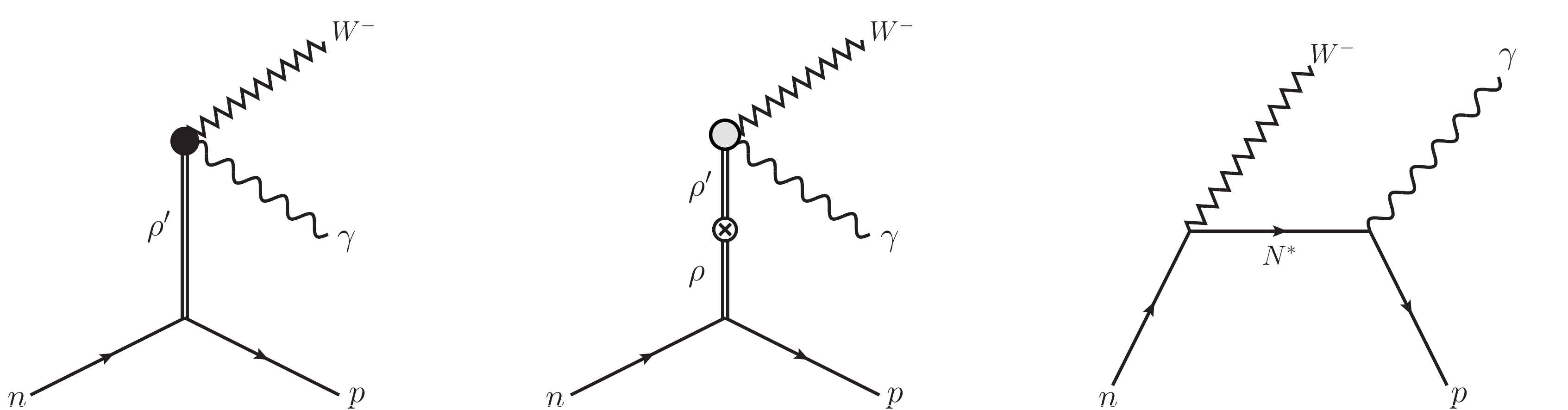}
\caption {Processes which could give rise to the $c_5$-dependent 
interaction of Eq.~(\ref{c5}). We use ``$N^*$'' to denote a 
nucleon resonance, 
and ``$\otimes$'' for $\rho-\rho^\prime$  mixing.}
\label{fig:sch}      
\end{figure}
{\it Interpreting a Limit on the T-odd Asymmetry---} 
The T-odd asymmetry is sensitive to the product Im($g_V c_5$). 
The value of Im($g_V$) can be bounded from the deviation of the empirical 
CKM unitarity test, namely, 
$|V_{ud}|^2+|V_{us}|^2+|V_{ub}|^2=0.99995 \pm 0.00061$~\cite{pdg12} from unity, 
to yield 
$\mathrm{Im}(g_V)<0.024$ at 68\% CL, 
The limit from the $D$ term is much sharper, as we have noted: 
Im $g_V < 7\times 10^{-4}$ at 68\% CL~\cite{nistD12}; 
a measurement of the T-odd asymmetry would limit Im($c_5$).   
The $c_5$ coefficient of Eq.~(\ref{c5}) can be generated in different ways, 
and we illustrate some possibilities in Fig.~\ref{fig:sch}, which include mixing
with new degrees of freedom, such as a ``hidden sector'' $\rho'$, as well as possible 
complex phases associated with the production of known 
nucleon resonances, or $N^*$'s.
We now develop a rudimentary model in which the $\rho'$ helps 
mediate a difference in the radiative $n$ and $\bar n$ $\beta$ decay rates. 
The notion of a hidden sector of strongly coupled matter 
is of some standing~\cite{okun,bezi}, and has more recently been 
discussed in the context of models which
provide a 
common origin to baryons and dark matter~\cite{nussinov,sekhar}, 
though the mechanism need not be realized through strong dynamics~\cite{kaplan,zurek} ---
we note Ref.~\cite{hoorabi} for a recent review. Intriguing
astrophysical anomalies have prompted the study of hidden sector models 
which permit couplings
to SM leptons; specifically, 
the visible and hidden sectors are connected through the 
kinetic mixing of the gauge bosons of 
their respective U(1) symmetries, notably through a 
SM hypercharge U(1)$_Y$ 
portal~\cite{Holdom:1985ag,Essig:2009nc,ArkaniHamed:2008qn,Baumgart}. 
Constraints on long-range 
interactions between dark-matter particles are 
sufficiently severe~\cite{Spergel:1999mh,Ackerman:2008gi,Feng:2009mn} that 
in such models the dark gauge symmetries are also 
broken through some dark Higgs sector~\cite{Baumgart}. 
In this paper we follow a different path. 
We consider a non-Abelian portal, mediated, e.g., 
by heavy scalars $\Phi$ which transform under the adjoint representation of the group; 
such an interaction can also be realized through kinetic mixing, generalizing from 
Ref.~\cite{Baumgart}, through 
${\rm tr}(\Phi F_{\mu\nu}){\rm tr}(\tilde\Phi {\tilde F}^{\mu\nu})$, as well as 
$\epsilon^{\mu\nu\rho\sigma} {\rm tr}(\Phi F_{\mu\nu})
{\rm tr}(\tilde\Phi {\tilde F}_{\rho\sigma})$, where 
$F^{a\,\mu\nu}$ is the SM SU(3)$_c$ field strength, and 
$\tilde{\Phi}^a$ and $\tilde{F}^{a\,\mu\nu}$ are 
fields and field strengths of a hidden strongly-coupled sector, nominally 
based on SU(3)$_{\tilde c}$. We anticipate 
that the dark matter candidate is a color singlet, so
that 
there are no dark long-range forces to negate. 
The connector is not a marginal operator, 
but the appearance of QCD-like couplings should make it more important in the 
infrared. 
To build a pertinent model at low energies we recall the hidden local symmetry
model of QCD~\cite{Bando,Kitano:2011zk}, in which the $\rho$ mesons function
as effective gauge bosons of the strong interaction. Upon including electromagnetism
this becomes a vector-meson dominance model, 
noting ``VMD1'' of Ref.~\cite{O'Connell:1995wf}, which we adapt to this case as 
\begin{equation} 
{\cal L}_{\rm mix} = -\frac{1}{4} \rho^a_{\mu\nu} \rho^{a\,\mu\nu}  - 
\frac{1}{4} {\rho^{\prime\,a}_{\mu\nu}} \rho^{\prime\,a\,\mu\nu}  + 
\frac{\epsilon }{2} {\rho^a_{\mu\nu}} \rho^{\prime\,a\,\mu\nu}  + 
\frac{m_\rho^2}{2} \rho^a_\mu \rho^{a\,\mu} + \frac{m_{\rho^\prime}^2}{2} 
\rho^{\prime\,a}_\mu \rho^{\prime\,a\,\mu} 
+ g_\rho J^{\mu\,a} \rho^a_\mu
\end{equation} 
where $J^{a\,\mu}$ denotes the baryon vector current and 
$\rho^{(\prime)\,a}$ are the gauge bosons of a hidden local SU(2) symmetry --- though 
$\rho^{(\prime)\,a}_{\mu \nu} = \partial_\mu \rho^{(\prime)\,a}_\nu - 
\partial_\nu \rho^{(\prime)\,a}_\mu$~\cite{O'Connell:1995wf}. 
Our model resembles those in 
Refs.~\cite{Holdom:1985ag,Essig:2009nc,ArkaniHamed:2008qn,Baumgart} but 
contains two massive vector fields. 
With $J_\mu^\pm = J_\mu^1 \pm i J_\mu^2$ and
$\rho^\pm_\mu = (\rho^1_\mu \mp i \rho^2_\mu)/\sqrt{2}$, the charged current 
pieces, dropping the mass terms, become 
\begin{equation}
{\cal L}_{mix}^\pm = -\frac{1}{4} \rho^{+\,\mu\nu} \rho^-_{\mu\nu}
-\frac{1}{4} \rho^{\prime +\,\mu\nu} \rho^{\prime\, -}_{\mu\nu}
+ \frac{\epsilon}{2} \left( \rho^{+\,\mu\nu} \rho^{\prime\, -}_{\mu\nu}
+ \rho^{-\,\mu\nu} \rho^{\prime\, +}_{\mu\nu}  \right) +
\frac{g_\rho}{\sqrt{2}} (\rho^+_\mu J^{+\,\mu} + \rho^-_\mu J^{-\,\mu})
\,.
\end{equation} 
The kinetic mixing term can be removed
through the field redefinition $\tilde \rho^\pm_\mu = \rho^\pm_\mu - \epsilon 
\rho^{\prime\,\pm}_\mu$, thus yielding a coupling of
the baryon vector current to $\rho^\prime$, as illustrated in the first panel 
of Fig.~\ref{fig:sch}, mimicking the role of the ``dark photon'' in fixed
target experiments~\cite{Bjorken:2009mm}. 
The $\rho^{\prime\,\pm}$ does not couple to photons; 
indeed, the particles of the hidden sector couple only to strongly interacting
particles --- we refer to Ref.~\cite{hoorabi} for discussion of models with
generalized conserved charges. We consider $m_{q} \sim {\cal O}(m_{q'})$ but with 
confinement scales $\Lambda^\prime < \Lambda$ so that $m_{\rho^\prime} < m_{\rho}$, noting
that dark and baryonic matter can have a common origin even if 
the dark matter candidate is lighter than the proton in mass~\cite{Buckley:2010ui}. 
Unlike related ``quirk'' models~\cite{Kang:2008ea}, 
the collider signatures of our scenario are minimal and are hidden within
hadronization uncertainties. However, if $m_{\rho^\prime} \lesssim 1\,{\rm MeV}$ 
it can be constrained by other low-energy experiments and observations; e.g., 
it can appear 
as a mismatch in the value of the neutron lifetime inferred from counting surviving 
neutrons from that inferred from counting SM decay products. 
It is also possible to build a model with additional hidden-sector 
portals. 
With a U(1)$_Y$ portal, e.g., the hidden quarks are allowed to have a milli-electric charge
if the dark-matter particle is an electrically neutral composite~\cite{Cline}. 
This possibility is illustrated in the ``mixed basis'' in the 
central panel of Fig.~\ref{fig:sch}. Limits on the SU(2)$_L$ and U(1)$_{em}$ couplings
follow, e.g., from studies of the $W^\pm$ width and the running of $\alpha$ and 
are significant; for 
simplicity we set this possibility aside. Thus limits on the 
T-odd asymmetry, for 
which a statistical error of ${\cal O}(10^{-3})$ could be achievable~\cite{NGH}, 
limits ${\rm Im}(c_5/M^2)= 
2\epsilon \,{\rm Im} \epsilon \,g_{\rho^0}^2/(16\pi^2 m_{\rho^\prime}^2)$
with $g_{\rho^0}\sim 3.3$~\cite{elster}. 

{\it SM Background---}
CP-conserving FSI in the SM can
induce T-odd decay correlations~\cite{callan,khrip}. 
A triple momentum correlation has been previously studied 
in $K^+\to \pi^0 l^+ \nu_l \gamma$ decay~\cite{braguta,khripK}, for which 
 both electromagnetic and strong 
radiative corrections enter, but the electromagnetic FSI 
effects are orders of magnitude larger~\cite{mueller}.
The small energy release associated with neutron 
and nuclear radiative $\beta$ decay imply that only electromagnetic radiative 
corrections can mimic the T-odd effect. 
The induced T-odd effects in this case have never been studied before, 
and we describe our calculation in Refs.~\cite{svgdh,svgdhA}. We neglect effects
of recoil order, which incurs corrections of ${\cal O}(Q_{EC}/M)$, and the nuclear 
computations are realized in the impulse approximation, so that the
effect of meson-exchange currents, estimated to yield corrections of 
${\cal O}(5-10\%)$~\cite{Chemtob:1971pu}, have been neglected. 
Some numerical results are shown in Table~\ref{fsidat}. 

\begin{table}[t]
\caption {Asymmetries from SM FSI in various weak decays. 
The range of the opening angle between the 
outgoing electron and photon is chosen to be  
$-0.9<\mathrm{cos}(\theta_{e\gamma})<0.9$.
}    
\label{fsidat}
\centering   \footnotesize
\vskip 0.5\baselineskip
\begin{center}
\begin{tabular}{l*{16}{c}r}
$\omega_{\rm min}({\rm MeV})$ && ${\cal A}^{\mathrm{FSI}}(n)$ && ${\cal A}^{\mathrm{FSI}}(^{19}\mathrm{Ne})$ && ${\cal A}^{\mathrm{FSI}}(^{35}\mathrm{Ar})$\\
\hline
0.01 && $1.76\times10^{-5}$  && $-2.86\times10^{-5}$ && $-8.35\times10^{-4}$\\
0.05 && $3.86\times10^{-5}$  && $-4.76\times10^{-5}$  && $-1.26\times10^{-3}$\\
0.1 && $6.07\times10^{-5}$   && $-6.40\times10^{-5}$ && $-1.60\times10^{-3}$\\
0.3 && $1.31\times10^{-4}$ && $-1.14\times10^{-4}$ && $-2.55\times10^{-3}$\\
\hline
\end{tabular}
\end{center}
\vspace{1ex} 
\end{table} 

The asymmetry from SM FSI is controlled by $(1-\rho^2/3)/(1+\rho^2)$~\cite{svgdh}, 
so that 
the best choice of nuclear
target 
is determined by noting that 
(i) the ${\rm Im}\, c_5$-induced T-odd asymmetry is bigger when the total energy release 
is bigger and (ii) the FSI effects can be suppressed if 
$\rho \sim 1.7$ and the daughter nucleus $Z$ is not too large.
Thus lighter nuclear candidates with allowed transitions of 
large energy release and $\rho \sim 1.7$ are the most useful for the BSM studies we suggest. 
Furthermore, the radiative $\beta$ decay branching ratio grows as 
the total energy release grows, making it easier to accrue statistics. 

{\it Summary--} The radiative $\beta$ decay of neutrons
and nuclei admits the study of a triple-product correlation in the decay-product momenta. 
This decay correlation is both parity and motion-reversal odd but spin-independent, 
making it sensitive to sources of CP violation beyond the SM which are not constrained
by searches for permanent EDMs. The appearance of the correlation is controlled by 
$\mathrm{Im}\,(g_Vc_5)$, where $c_5$ is the low-energy constant of the 
pseudo-Chern-Simons operator 
first noted by Harvey, Hill, and Hill~\cite{HHHprl,HHHprd,Hill2010}. 
The size of the SM background is small relative to the anticipated experimental sensitivity. 
Empirical limits on the triple-product correlation can 
be interpreted as limits on the CP-violating kinetic mixing of the gauge bosons of
QCD with a strongly coupled hidden sector, possibly giving new insights on 
the origin of baryons and dark matter. 

We acknowledge partial support from the U.S. Department of Energy under
contract no.\ DE-FG02-96ER40989, and 
we thank Oscar Naviliat-Cuncic and Jeffrey Nico for information regarding
experimental aspects of radiative $\beta$ decay.

\end{document}